\documentclass[a4paper]{jpconf} 
\usepackage{graphicx}
\begin{document}
\newcommand{\ignore}[1]{}
\title{Predicting entanglement and coherent times in FMO complex using the HEOM method}
\author{Bruno Gonz\'alez-Soria, Francisco Delgado and Alan Anaya-Morales}
\address{Tecnologico de Monterrey, School of Engineering and Sciences, Mexico}
\ead{fdelgado@tec.mx}

\begin{abstract}
Fenna-Matthews-Olson (FMO) bacteriochlorophylls (BChls) are molecules responsible of the high efficiency energy transfer in the photosynthetic process of green sulfur bacteria, controversially associated to quantum phenomena of long lived coherence. This phenomenon is modelled using Quantum Open Systems (QOS) without included memory effects of the surrounding approximated as a phonon bath on thermal equilibrium. This work applies the Hierarchical Equations of Motion method (HEOM), a non-Markovian approach, in the modelling of the system evolution of the FMO complex to perform predictions about the coherence time scales together with global and semi-local entanglement during the quantum excitation.
\end{abstract}

\section{Introduction: FMO complexes and their quantum dynamics simulation}

\begin{figure}[t] 
\begin{center}
\setlength{\abovecaptionskip}{1pt}
\includegraphics[width=30pc]{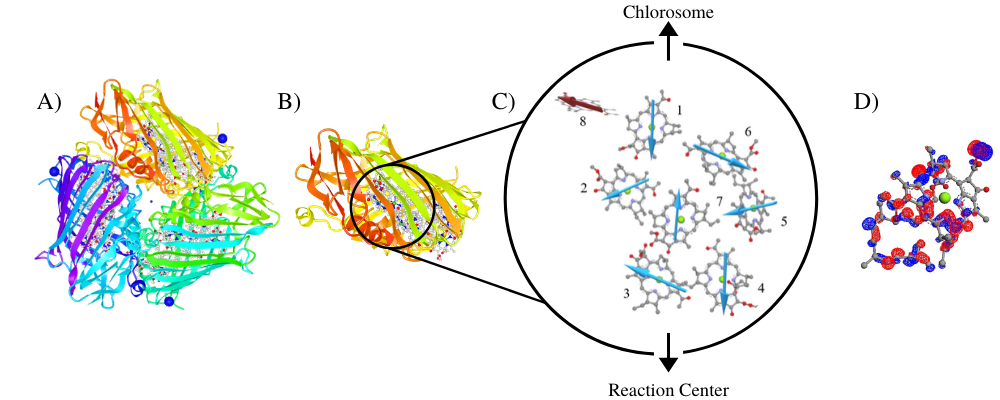}
\caption{FMO from {\it Prosthecochloris aestuarii}. A) Complete structure of the protein trimer, B) close-up to a single monomer, C) the eight numbered chromophores inside with their dipole momenta and D) a single BChl $a$ molecule. Figure produced from Protein Data Bank file 3EOJ.}\label{FMO_BChla_dipoles}
\end{center}
\end{figure}

Photosynthetic bacteria have evolved to transform sun energy into biochemical energy through  physicochemical mechanisms carried out in specialized chemical structures, the FMO complex (Figure \ref{FMO_BChla_dipoles}), a protein structure responsible of energy transfer with a nearly 100\% efficiency from the Light Harvesting Antennas (LHA) to the reaction center (RC) in green sulphur bacteria \cite{fmo2}. Figure \ref{FMO_BChla_dipoles} shows the FMO structure with the eight inner bacterioclorophylls (BChls), their dipole momenta and its scaffolding protein structure. Ultrafast spectroscopic studies reveal long time quantum coherence between the electronic states of the BChls \cite{Engel}, a mechanism of high transfer efficiency sampling the energy space through the excitonic superposition. 

QOS theory has been applied to study such phenomena in the quantum arena through the density matrix characterizing the energy statistical ensemble. The most common approaches, Redfield and Lindblad master equations \cite{Sarovar, Jesenko} do not consider the non-markovian behaviour of the structure protein vibrations, modeled as a phonon bath interacting with the BChls. A more realistic model should consider the bath relaxation, to understand the chromophore-protein interaction role on the energy transfer and the tunning effect on the site energies.
 
Current work reproduces the excitation energy transfer by the eight BChls within each monomer of the FMO complex using the HEOM method. First section discusses the Hamiltonian for the system and HEOM \cite{Tanimura}. Second section provides an analysis of coherence and entanglement achieved in the dynamics. Conclusions discusses an insight about the possible relation between the parametric effects on the population dynamics and the genetic traits of such complexes.

\section{Modelling the quantum dynamics inside a monomer of an FMO complex}

Modelling of FMO complexes departs from a Hamiltonian $H_S$ reproducing the excitations due to dipole-dipole interactions among BChls (the {\it system}, $S$). Experimental analysis shows each ${\rm BChl}$ becomes at most excited to the first energy level \cite{Weidemuller} and only one at the time, then the Hilbert space is spanned by the excitation state obtained from the tensor product of $\vert 0_i \rangle$ (ground state) and $\vert 1_i \rangle$ (first excited state) for the entire system of $N$ BChls ($N=7, 8$ depending on the considered model): $\vert k \rangle \equiv \vert 0_1 0_2 ... 1_k ...0_N \rangle$, the {\it occupation} basis. While, the protein monomer where the  BChls are embedded works as a scaffold. Since a protein has a large number of atoms compared to other molecules,  vibrational states could be considered on the continuous regime: a phononic medium (the {\it bath}, $B$) exchanging energy with the set of BChls. The whole Hamiltonian and the density matrix $\rho_T$ ($S$ plus $B$) fulfils the von Neumann-Liouville equation:

\begin{eqnarray}
    H_T &=& H_S + H_{\rm reloc} + H_B + H_{S-B} \quad \rightarrow \quad {\dot \rho_T} = - \frac{i}{\hbar} [H_T,\rho_T] \\
    \quad && {\rm with:}  \quad H_S = \sum_{i=1}^N E_i \vert i \rangle \langle i \vert + \sum_{1 \le i,j \le N} J_{ij} (\vert i \rangle \langle j \vert + \vert j \rangle \langle i \vert) 
\end{eqnarray}

\noindent $E_i$ and $J_{i}$ are reported by \cite{Adolphs,Schmidt} for $N=7, 8$ respectively. Developed by \cite{Tanimura} for phononic media and then applied to FMO by \cite{Fleming}, HEOM follows by switching into the interaction picture of $H_{S-B}$ in order to trace the bath system considering non-markovian considerations. It becomes in recursive equations considering $D$ previous temporal stages of the bath labeled by a vector ${\bf n}$ of degree $s=0,1,...,D$ and $N$ for the BChl considered. There, $\rho_{\bf n}$ with ${\bf n}=(0,..,0)$ is the density matrix of the system and other $\rho_{\bf n}$ auxiliary ones corresponding to all vectors ${\bf n}=({\rm n}_1,{\rm n}_2,...,{\rm n}_N)$ with ${0 \le \rm n}_k \in {\rm Z}^+ \cup \{0\}$ such as $\sum_{k=1}^N {\rm n}_k = s$. If $k$ is the Boltzman constant, $\beta=k T$ and  $V_k=\vert k \rangle \langle k \vert$. $\Delta_k={\lambda_k}/{\beta \hbar^2 \gamma_k}$ (a reorganization term appearing in QOS due to the interaction with the bath). $\gamma_i$ comprises the interaction strengths between the bath and each BChl coming from a bilinear model of system and bath operators for $H_{S-B}$. HEOM model is written as:

\begin{eqnarray} \label{HEOM}
\dot{\rho_{\bf n}} &=& -\frac{i}{\hbar} [H_{S}+ \sum_{k=1}^N \lambda_k V_k,\rho_{\bf n}] - \sum_{k=1}^N  {\rm n}_k \gamma_k \rho_{\bf n} - r_{\rm trap} (\{ V_3, \rho_{\bf n} \} + \{ V_4, \rho_{\bf n} \}) \nonumber \\ 
&& \quad + i \sum_{k=1}^N\Delta_k (  [V_k,\rho_{{\bf n}_{k+}}] + {\rm n}_k ([V_k,\rho_{{\bf n}_{k-}}] - i \frac{\beta \hbar \gamma_k}{2} \{V_k,\rho_{{\bf n}_{k-}}\}))
\end{eqnarray} 

\noindent  BChls 1, 6 and 8, work as FMO antennas while BChls 3 and 4 drive the energy oscillations to the RC at the trapping rate, $r_{\rm trap}$. If ${\bf n}$ is a vector of order $s$, then ${\bf n}_{k \pm}$ is the vector of order $s\pm 1$ obtained from ${\bf n}$ by increasing (decreasing) its component ${\rm n}_k$ by one. Last model has been considered for $N=7$ to model the FMO complex dynamics \cite{Kreisbeck}. It is computationally convenient translate this equation to the superoperator-supervector version \cite{Gonzalez1}. Next section solves numerically HEOM model for $N=8$ to simulate the dynamics, then analyzing the coherence and some global or bipartite entanglement measures among FMO complex BChls.

\section{Coherence times and entanglement for the FMO complex}

BChls exhibit an increase and holding of quantum coherence, in raw terms, the prevalence of terms $\rho_{ij}, i \ne j$ in $\rho$. A measure based on the minimum distance from any non-coherent state:

\begin{eqnarray}\label{cl1}
C_{l_1}(\rho) =\min_{\sigma \in \Gamma} \sum_{ij} \vert \rho_{ij} - \sigma_{ij}\vert = 2 \sum_{i<j}  \vert \rho_{ij} \vert \in [0,d-1]
\end{eqnarray}

\noindent the $l_1-$norm \cite{horn}. $\Gamma$ is the set of non-coherent states, with the minimum reached for $\sigma_{\rm diag}=\sum_i \rho_{ii} \vert \phi_i \rangle \langle \phi_i \vert$. The maximum bound is for $\vert \psi \rangle = \frac{1}{\sqrt{d}} \sum_{i=1}^d \vert i \rangle$. For the entanglement, we use the concurrences running from zero (separable) to one (maximally entangled), obtained by partially tracing $\rho$ except for BChls $k,l$, $\rho_{kl}={\rm Tr}_{\{kl\}}(\rho)$ or BChl $k$, $\rho_{\{k\}}={\rm Tr}_{\{k\}}(\rho)$: 

\begin{eqnarray}
{\mathcal C}_{\{kl\}} &\equiv& {\mathcal C}({\rho}_{\{kl\}}) = 2 \vert \rho_{kl} \vert \in [0,1] \\
{\mathcal C}_{\{k\}} &\equiv& {\mathcal C}({\rho}_{\{k\}}) = 2 \sqrt{\rho_{kk} (1 - \rho_{kk})} \in [0,1]
\end{eqnarray}

\noindent They are respectively interpreted as the entanglement a) among systems $k, l$ \cite{Sarovar}, and b) between a system $k$ and the remainder \cite{Gonzalez1}. Note each pair entanglement contributes to $C_{l_1}$ (\ref{cl1}). 

\begin{figure}[t] 
\vspace*{-1.5cm} 
\begin{center}
\begin{tabular}{c c}
      \includegraphics[width=77mm]{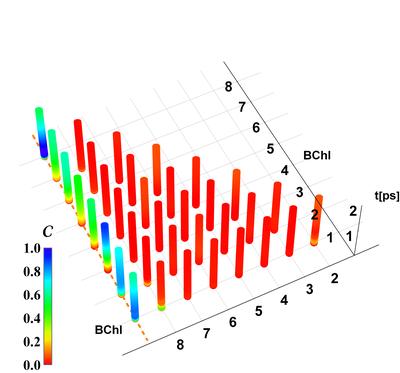} &  \includegraphics[width=77mm]{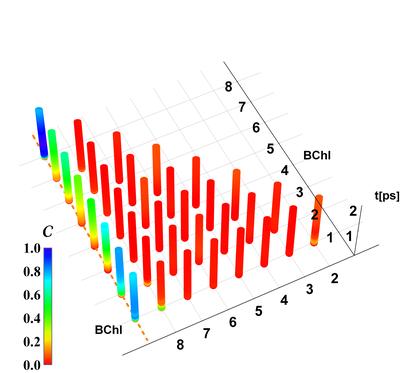} \\[0.3cm]
      \includegraphics[width=77mm]{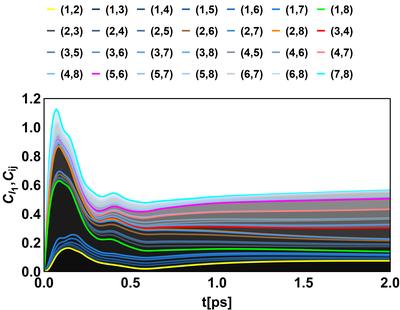} &  \includegraphics[width=77mm]{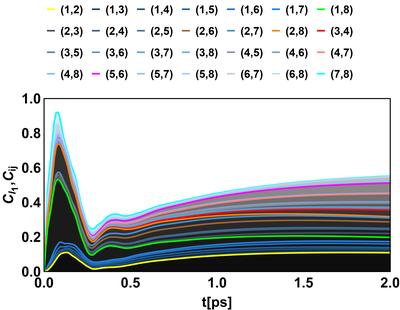} \\[0.4cm]
     \bfseries{{a) $\lambda_k = 35{\rm cm}^{-1}$}} & \bfseries{{b) $\lambda_k = 65{\rm cm}^{-1}$}}
\end{tabular}
\caption{For a) $\lambda_k=35{\rm cm}^{-1}$ and b) $\lambda_k=65{\rm cm}^{-1}$ : Time evolution in $0-2$ ps in the vertical axis for ${\mathcal C}_{\{kl\}}$ among BChl pairs and ${\mathcal C}_{\{k\}}$ of each BChl with the remainder monomer in color (up). $C_{l_1}(\rho)$ coherence and each ${\mathcal C}_{\{kl\}}$ cumulative contribution of the pairs by layers (down).}\label{fig2}
\end{center}
\end{figure}

Thus, by solving the dynamics for one monomer of the FMO complex using the HEOM method for $N=8$ BChls and $D=3$ at room temperature $T=293^{\circ} K$, we trace the entanglement measures ${\mathcal C}_{\{kl\}}$, ${\mathcal C}_{\{k\}}$, and then the concurrence $C_{l_1}(\rho)$ during the evolution in the interval time $0-15$ ps. There, the same characteristic parameters are used for all BChls: $\gamma_k=50 {\rm cm}^{-1}$, $r_{\rm trap}=1 {\rm ps}^{-1}$ and two reorganization energy values $\lambda_k=35 {\rm cm}^{-1}, 65 {\rm cm}^{-1}$ \cite{Sarovar} (as $H_{S}$, in the spectroscopic units, transformed to ${\rm cm}^{-1}$ dividing by $200 \pi \hbar c$; ${\rm cm}^{-1}$ to ${\rm s}^{-1}$ with factor $200 \pi c$). Outcomes are reported in the Figure \ref{fig2} for $0-2$ ps, the more meaningful interval for the analysis.

Figures \ref{fig2}a-b exhibits for $\lambda_k=35 {\rm cm}^{-1}$ and $\lambda_k=65 {\rm cm}^{-1}$ respectively, upward, the entanglement ${\mathcal C}_{\{kl\}}$ by pairs evolving vertically in agreement with the color bar besides. BChl 8 is assumed initially excited. Note the sudden strong entanglement among BChls 1, 2 and 8 which turn off before 0.5 ps. Together, in the first row, we report ${\mathcal C}_{\{k\}}$ for each BChl. Note a similar behavior but extended in time, reflecting the entanglement transference to the remaining BChls. Downward, the $C_{l_1}(\rho)$ coherence reported as the upper contour (cyan), showing its additive components ${\mathcal C}_{\{kl\}}$ (in cumulative layers) for each BChls pair. The main initial contributions are due to the pairs $(1,2), (1,8)$ and $(2,8)$. The end is ruled by the coherences among pairs $(3,4), (4,7), (2,6)$ and $(5,6)$ in outstanding colors. It is true for both values of $\lambda_k$, despite the time scale slows for $65 {\rm cm}^{-1}$. Note the intermediate dropping of the coherence around $t=0.25 {\rm ps}$ deeper for $\lambda_k=65 {\rm cm}^{-1}$ setting a landmark for the decoherence times.

\section{Conclusions}

HEOM method was used to model the dynamic evolution of the eight BChls in one monomer of the FMO complex at room temperature depicting the boost of long-live coherences via localized entanglement, first among BChls 1, 2, 8 and then transferred to the remainder BChls conducting to the final populations in BChls 3 and 4 exiting the energy to the RC. Reorganization energy $\lambda_k$ values suggest an important role in the time scale and the behavior of such coherence phenomena possibly related with the strains efficiency exhibited in their spectroscopic characterization.

\section*{Acknowledgements}
The authors would like to acknowledge the financial support of NOVUS 2019 PHHT023-19ZZ00018, an initiative of Tecnologico de Monterrey, Mexico, in the production of this work.

\end{document}